# The science enabled by a dedicated solar system space telescope

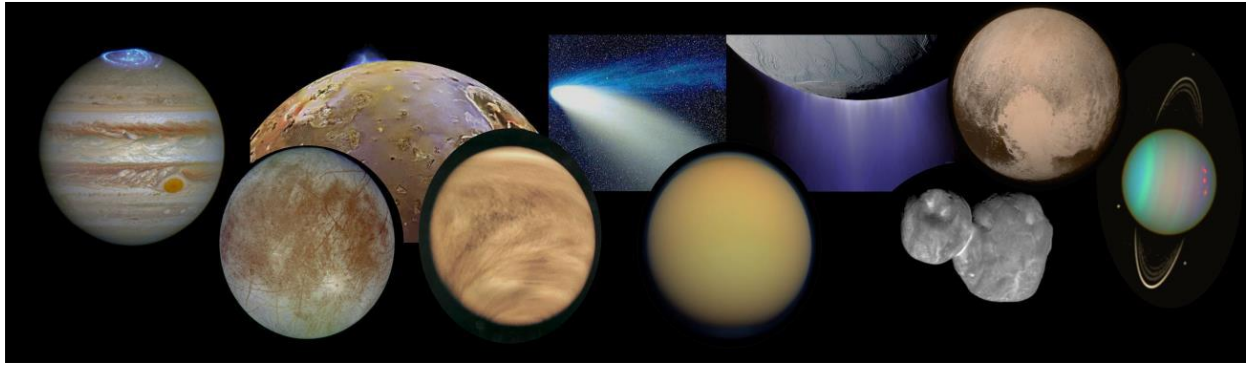

[1]C.L. Young, [2,21]M.H. Wong, [3]K.M. Sayanagi, [2]S. Curry, [4]K.L. Jessup, [5]T. Becker, [6]A. Hendrix, [7]N. Chanover, [8]S. Milam, [9]B.J. Holler, [10]G. Holsclaw, [11]J. Peralta, [12]J. Clarke, [4]J. Spencer, [13]M.S.P. Kelley, [2]J. Luhmann, [1]D. MacDonnell, [14]R.J. Vervack Jr., [5]K. Retherford, [15]L.N. Fletcher, [2]I. de Pater, [6]F. Vilas, [13]L. Feaga, [2]O. Siegmund, [16]J. Bell, [17]G. Delory, [17]J. Pitman, [5]T. Greathouse, [2]E. Wishnow, [18]N. Schneider, [2]R. Lillis, [19]J. Colwell, [1]L. Bowman, [20]R.M.C. Lopes, [21]M. McGrath, [21]F. Marchis, [21]R. Cartwright, [5]M.J. Poston

Corresponding author: Cindy L. Young; cindy.l.young@nasa.gov

[1]NASA Langley Research Center, Hampton, VA
[2]University of California, Berkeley, CA
[3]Hampton University, Hampton, VA
[4]Southwest Research Institute, Boulder, CO
[5]Southwest Research Institute, San Antonio, TX
[6]Planetary Science Institute, Tucson, AZ
[7]New Mexico State University, Las Cruces, NM
[8]NASA Goddard Space Flight Center, Greenbelt, MD
[9]Space Telescope Science Institute, Baltimore, MD
[10]University of Colorado, Boulder, CO
[11]Japan Aerospace Exploration Agency, Sagamihara, Japan
[12]Boston University, Boston, MA
[13]University of Maryland, College Park, MD
[14]Johns Hopkins Applied Physics Laboratory, Laurel, MD
[15]University of Leicester, Leicester, UK
[16]Arizona State University, Tempe, AZ
[17]Heliospace Corporation, Berkeley, CA
[18]University of Colorado Laboratory for Atmospheric and Space Physics, Boulder, CO
[19]University of Central Florida, Orlando, CA
[20] Jet Propulsion Laboratory, California Institute of Technology, Pasadena, CA
[21]SETI Institute, Mountain View, CA

# 1. Introduction and Motivation

The National Academy Committee on Astrobiology and Planetary Science (CAPS) made a recommendation to study a large/medium-class dedicated space telescope for planetary science, going beyond the Discovery-class dedicated planetary space telescope endorsed in *Visions and Voyages*[0]. Such a telescope would observe targets across the entire solar system, engaging a broad spectrum of the science community. It would ensure that the high-resolution, high-sensitivity observations of the solar system in visible and UV wavelengths revolutionized by the Hubble Space Telescope (HST) could be extended. A dedicated telescope for solar system science would a) transform our understanding of time-dependent phenomena in our solar system that cannot be studied currently under programs to observe and visit new targets and b) enable a comprehensive survey and spectral characterization of minor bodies across the solar system, which requires a large time allocation not supported by existing facilities. The time-domain phenomena to be explored are critically reliant on high spatial resolution UV-visible observations. The key questions identified in this whitepaper address the crosscutting themes of planetary science from planetary habitability, origin and evolution of the solar system, and understanding processes that drive present-day dynamics. This paper presents science themes and key questions that require a long-lasting space telescope dedicated to planetary science that can capture high-quality, consistent data at the required cadences that are free from the complicating effects of the terrestrial atmosphere and differences across observing facilities. Using astrophysical assets for solar system science is important, but cannot fully address the science described here, given the number and time-varying nature of solar system targets. For instance, just 6% of HST time has been allocated to solar system targets over the lifetime of the mission[1]. Such a telescope would have excellent synergy with astrophysical facilities by placing planetary discoveries made by astrophysics assets in temporal context, as well as triggering detailed follow-up observations using larger telescopes. The telescope would also support future missions to the Ice Giants, Ocean Worlds, and minor bodies across the solar system by placing the results of such targeted missions in the context of longer records of temporal activities and larger sample populations.

## 2.1. Active Plumes and Volcanism

Geologic activity may be key to understanding the past/present habitability of ocean worlds and Venus[2]. Given the likely sporadic nature of plume events that may be observed by *Europa Clipper* and *JUICE* in the late 2020s, a long-duration observing campaign would complement spacecraft observations by providing the temporal context to enable interpretation of trends in the data. This theme encompasses three key questions:

*[1] Are Venus and Titan volcanically active today?* The telescope could search for activity on Titan and Venus, where the question of active volcanism has been debated for years[3], spectroscopically seeking anomalies in surface emissivity (Fig. 1A)[4] and $H_2O$ concentration in the near-infrared (NIR)[5], while imaging in the UV to observe potential bright spots in the clouds[6]. Given sufficient resolution, the telescope could also look for signs of cryovolcanic water vapor on icy bodies beyond Saturn (e.g, Miranda, Ariel, Triton, and Pluto). *[2] What drives variability in volcanic and cryovolcanic activity?* Periodicities of activity revealed by such a telescope may determine forces driving Europa's putative plumes (Fig. 1B)[7]: tidal stresses as on Enceladus vs. sporadic filling of reservoirs as on Earth and presumably Io. Eruption statistics would include measurements of changing surface deposits (e.g., surface salts[8], or silicate ash) and observations at UV wavelengths to allow for detection of limb plume eruptions via water emissions[7], scattering of ice/dust particles, or gas-phase absorptions during satellite transits of the disk of Jupiter/Saturn[9,10]. *[3] What is the composition of magma and cryomagma reservoirs?* By

observing lava flows on Io before and while they cool, the telescope could distinguish basaltic or ultramafic compositions[11], constraining models of Io's interior.

*Science requirements and trades:* Icy satellite water plume detections (including O and H) requires UV spectroscopy from 120-140 nm. Observations of the SO 1.7 µm emission band on Io during eclipse provide a way to study the activity of stealth volcanoes, which are enigmatic and may be widespread on Io[12, 13]. Venus and Titan surface/lower atmosphere investigations require spectroscopy in NIR windows spread over 0.9-1.7 µm. Additionally, Venus nightside surface emissivity observations require solar elongation limits of 30° or less, placing requirements on sunshield design. A 2-m telescope can deliver the required UV resolution of ≤120 km for plumes in the Jovian and Saturnian systems, and a VNIR resolution of ≤200 km for surface emissivity anomalies on Venus. Characterizing variability of surface deposits, lava flows, thermal anomalies, and surface emissivities in the outer solar system, as well as searching for new cryovolcanic activity beyond Saturn, and studying stealth volcano activity on Io, require the superior resolution afforded by a 10 m telescope.

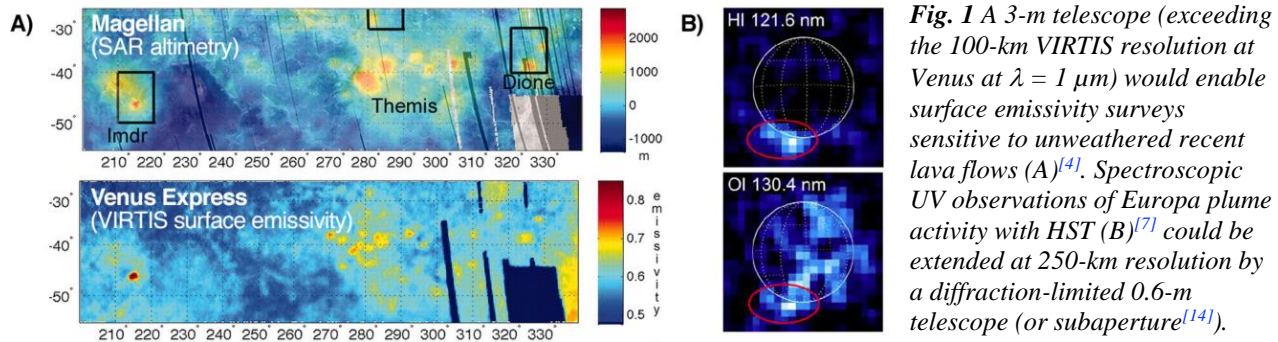

*Fig. 1* A 3-m telescope (exceeding the 100-km VIRTIS resolution at Venus at $\lambda = 1$ µm) would enable surface emissivity surveys sensitive to unweathered recent lava flows (A)[4]. Spectroscopic UV observations of Europa plume activity with HST (B)[7] could be extended at 250-km resolution by a diffraction-limited 0.6-m telescope (or subaperture[14]).

## 2.2. Outer Solar System Minor Body and Irregular Satellite Survey

Remote spectroscopy across the UV, visible, and NIR spectral ranges of Jupiter Trojans, irregular satellites, Centaurs, and Kuiper Belt Objects (KBOs) provide the most efficient means of probing the initial composition of the solar nebula and the process of planetary migration.

*[1] What do the compositions/colors of minor bodies/irregular satellites reveal about planetary migration early in solar system history?* Models and observations suggest the irregular satellites[15] and Jupiter Trojans[16] originated in the Kuiper Belt. Spectral characterization across these populations would establish links between current and primordial populations, informing models of giant planet migration[17]; time allocation of existing facilities does not support complete surveys, and space observations reach higher background-limited magnitudes. *[2] What dynamical processes are shaping minor body populations today?* Scattered Disk Objects (SDOs) in the Kuiper Belt interact gravitationally with Neptune and may be the source population of some Centaurs[18]. Those Centaurs present an opportunity to study KBOs in detail and understand how their surfaces evolve as they migrate toward the Sun[19]. UV spectral characterization of surface space weathering[20] could definitively link SDOs and Centaurs. *[3] What do the compositions of minor bodies reveal about the radial variations in the solar nebula?* Radial gradients of ice abundances in the solar nebula are a suggested cause of KBO visible colors[21]. KBO surfaces become increasingly water ice-rich at absolute magnitudes (HV) of 3-6, corresponding to diameters of ~300-1200 km[22]. This transition (Fig. 2) marks the size limit for differentiation and is tied to formation distance and interior composition.

*Science requirements and trades:* Spectra (0.1-1.7 µm; R~50) of ~1000 objects would include ~200 Trojans, all observable irregular satellites, 400 KBOs from all dynamical classes, all observable Centaurs, a comparable number of SDOs, and ~100 KBOs within the HV range of 3-6. Assuming a limiting *V* magnitude of 24, this minimum overall sample would be observed in 1-4 months (assuming 1-3 hours/target). Trade studies will determine the required stability and spectroscopic aperture sizes, which affect the accurate placement of moving targets.

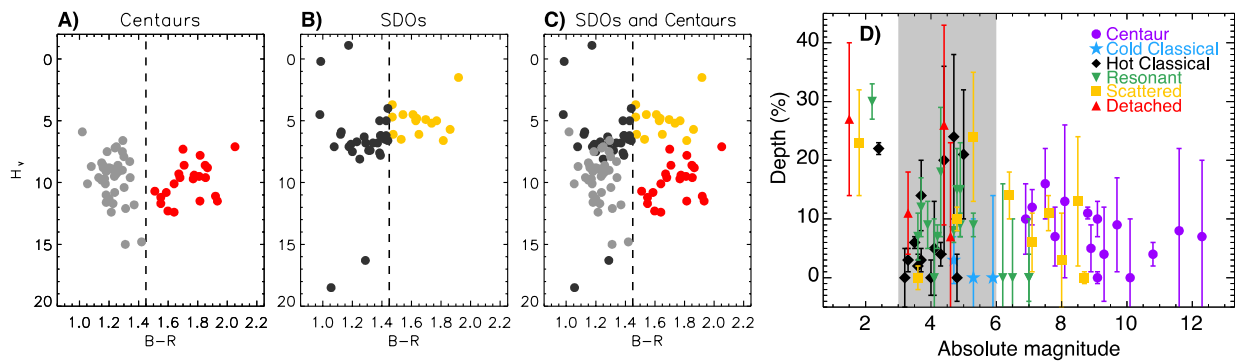

***Fig. 2*** *Broadband color data[23] (for (a) Centaurs, (b) Scattered Disk Objects (SDOs), and (c) both overplotted) cannot conclusively validate the dynamically-based hypothesis that Centaurs originate from the SDOs, requiring a spectroscopic sample from each population. (d) The transition region from water-rich to water-poor surfaces is shown in grey, in a plot of water ice feature strength vs. absolute magnitude[22]. The telescope would increase the sample size of KBOs and Centaurs by an order of magnitude and enable comparisons of water ice abundance, color, and diameter, shedding light on the formation and evolution of the various dynamical populations.*

### 2.3. Dynamic Atmospheres

Evolving processes in solar system atmospheres can be compared with those observed in Earth's atmosphere and ocean[24] as well as exoplanets and Brown Dwarfs[25], where transit spectroscopy, light-curves, and Doppler imaging have demonstrated the diversity and time variability of clouds, hazes, and circulation regimes[26]. A space observatory targeting multiple planets enables a powerful comparative planetology approach toward understanding energy transport across the range of examples offered in our own solar system; such insights can be further applied to the thermal evolution of exoplanets[27].

***[1] How does energy/momentum transport vary temporally and spatially in dense planetary atmospheres?*** Long-duration imaging campaigns characterize the variability of dynamical features like convective storms, waves, vortices, and jets (Fig. 3), surpassing datasets like the *Cassini* Jupiter approach movie or the *Voyager* Neptune Great Dark Spot sequence[24,28]. Winds can be measured by tracking clouds in the atmospheres of Venus and giant planets. Airglow on Venus characterizes horizontal transport through dayside photodissociation and nightside recombination[25,29], and vertical energy transport via waves[30]. ***[2] How is vertical energy transport modulated by chemical and thermodynamic processes?*** Latitudinally shifting hazes on Venus, Titan and the giant planets, and active photochemistry revealed in UV spectra of Venus, Jupiter and Saturn[31-33] reflect meridional overturning circulations. Convection shows seasonal and other periodicities potentially driven by solar forcing on Titan[34], a balance between internal heat release and thermodynamic convective inhibition on Jupiter[35], or a combination of both on Saturn. Clouds on Venus may also be influenced by the solar cycle[36]. ***[3] What is the current impactor flux and size distribution in the outer solar system?*** The largest impacts on Jupiter suggest a 20-fold discrepancy from the impactor flux predicted by comet dynamical models and craters on icy satellites[37]. The telescope will search for impact events to determine the impactor size distribution and may also discover impacts on Saturn, Uranus, and Neptune, where recent impacts are inferred but have not been witnessed[38-40].

***Science requirements and trades:*** A study is needed to prioritize spectral observations directly and during stellar occultations (< 230 nm, R>500), vs. multispectral imaging for particle properties, wind retrievals, and feature tracking (filters from 250 nm to 2 μm, driving requirements for long wavelength cutoff). Impacts could be detectable a few times per year[37] with sustained 5-hour cadence high-resolution observations in near-UV (NUV) and methane-band filters. Neptune's dark vortices (2000-km scale; never observed from the ground[41]) illustrate resolution requirements: these features can be tracked with a 1.2-m telescope (1800-km resolution), their sizes determined with a 2-m telescope (1100-km resolution), and shape oscillations (Fig. 3D) measured with a 10-m telescope (200-km resolution). Wind tracking may be achieved on Venus' dayside at solar elongations ~ 40 to 50°; but nightside wind tracking and airglow detections require solar elongation limits of 30° or less, placing requirements on sunshield design.

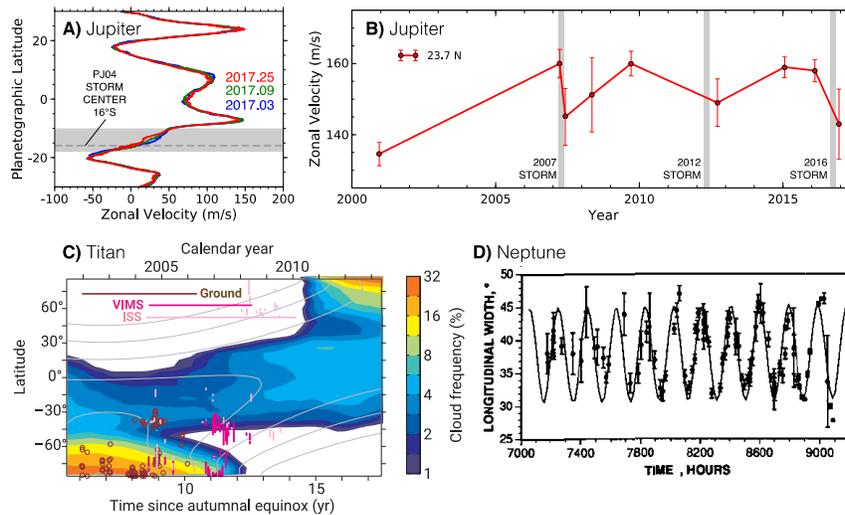

*Fig. 3 There are gaps in our understanding of storm/cloud activity, jets, and vortices of all planets with atmospheres due to the limited temporal coverage currently available. Major storm eruptions in Jupiter's southern (A) and northern (B) hemisphere alter zonal winds[42]. Models[34] duplicate storm activity at Titan's pole but not at mid-latitudes (C). Oscillations in the shape of Neptune's Great Dark Spot (D) from Voyager's Neptune approach give insights into deep stratification, wind shear, and chemistry[24].*

## 2.4. Magnetospheric Interactions

The outer planets provide natural laboratories to understand how strong intrinsic magnetic fields interact with the solar wind and internal plasma sources. Auroras are produced at these planets when charged particles precipitate into the upper atmosphere and excite molecules into higher electronic states, which emit photons as they decay. The upper atmospheres of Mars and Venus, unprotected by intrinsic magnetic fields, interact directly with the solar wind.

***[1] What controls auroral processes on different timescales?*** On Jupiter and Saturn, the solar wind flows past the planet in hours to days[43] vs. a few minutes for Earth. Spectral mapping of auroral emission would provide critical observational information to establish the relationship between timescales and field geometries. The telescope could map variation in the faint UV auroral emissions at Uranus, Neptune, and some ocean worlds, along with several types of auroras at Mars and Venus[44,45]: diffuse auroras responding to energetic particles produced by solar flares or interplanetary shocks, and proton auroras discovered by MAVEN at Mars and expected (though not yet observed) at Venus. These UV auroras cannot be observed from the ground, so a space telescope is necessary. ***[2] What is the balance between internal/external control of magnetospheric variability?*** The telescope would provide a stable vantage point to observe multiple parts of the Jupiter/Saturn systems, characterizing variabilities of (1) sources like Io's volcanoes and atmosphere, (2) ion and neutral tori of Io and Enceladus, and (3) giant planet auroras, including source flux tube footprints. Unexplained variability in the Enceladus torus[46] could be tested for correlation with the moon's plume output's dependence on the orbital position[47]. Although Jupiter's solar wind response differs from Saturn's (Fig. 4), measurements of

solar wind density and speed can be extrapolated from 1 AU to Saturn and Jupiter, particularly during the 2-3 months of planetary alignment[48].

*Science requirements and trades:* Measurements consist of imaging and spectroscopy that target atomic/molecular/ion emissions of H, $H_2$, O, $CO_2$, and S in the UV/visible from 115-162 nm, 190-300 nm, and 558 nm. Covering large tori around Jupiter/Saturn requires large FOV or tiling. The Io flux tube footprint can be fully resolved by a 2-m telescope (diffraction limited at 200 nm) but not a 1.2-m telescope[49], while the Enceladus footprint (not resolved by Cassini UVIS[50]), could be resolved by a 10-m telescope (31 km diffraction limit at Saturn at 200 nm).

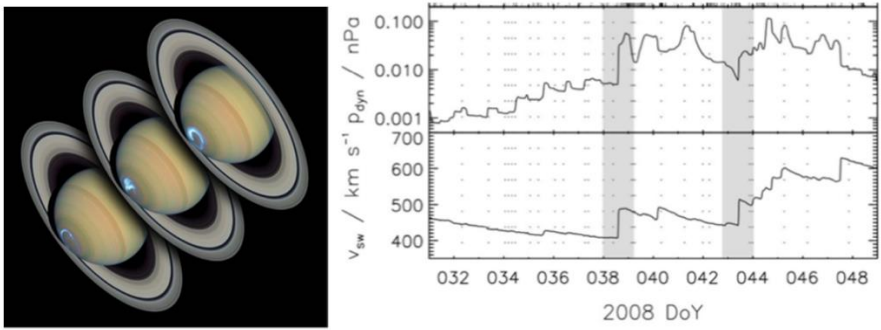

*Fig. 4* (left) HST far-UV images of Saturn's aurora and changes during an auroral storm, and (right) total auroral power at Saturn vs arriving solar wind speed. The shaded regions indicate the arrival of solar wind shocks at Saturn[51].

### 2.5. Planetary Ring Systems

The study of planetary ring systems is not only critical for understanding the dynamic history of our own solar system, but also sheds light on physical processes that lead to planet formation in protoplanetary disks. The telescope would address outstanding questions related to ring systems of Jupiter, Saturn, Uranus, Neptune, Chariklo, Haumea, and other potentially ringed minor bodies:

*[1] What are the current and past environments of planetary rings across the solar system?* Direct spectral measurements provide comparative studies of the composition and processing environments across planetary ring systems. Active satellites and micrometeorite impacts can affect the color of the rings and create a temporally variable ring[52]. Spectral observations of Chariklo have revealed icy rings around an otherwise dark object through observations at different ring opening angles throughout its orbit[53]. *[2] How do ring structures evolve and interact with nearby and embedded moons?* Stellar occultations from Earth provide observations of ring structures with spatial resolutions at 10s of kilometers[54]. Observations at small ring opening angles can better detect and characterize dusty rings[55] and self-gravity wakes[56], and those at large ring opening angles record shapes of ring edges and structures. Diffracted light signatures during occultations characterize the particle size distribution, tied to the dynamical interactions between ring particles and nearby moons[57]. UV diffraction signatures trace micron and millimeter-sized particles liberated by ongoing collisional activity in the rings[58], benefitting from reduced solar reflectance background.

*Science requirements and trades:* The spatial resolution retrieved by occultation light-curves is proportional to detector readout speed. Time-critical occultations occur ~1–4 times/year per object[59], with typical durations lasting <10 hours. Compositional variation affects spectral slopes in the visible and NUV, and $H_2O$ ice bands near 1.25, 1.6, and 1.65 μm. The far-UV is particularly sensitive to water ice with a strong absorption feature at 165 nm. Annual observations of reflected sunlight from the rings will attain spectral measurements at different ring opening angles. Telescope designs to minimize ill-effects of scattered planetary light must be considered.

## 2.6. Cometary Evolution, Morphology, and Processes

Many of the primary molecular emissions (e.g., from the parent molecules $H_2O$, $CO_2$, CO) and several of their daughter products are difficult or impossible to measure from the ground due to the terrestrial atmosphere (Fig. 5)[60] and transient behaviors like outbursts cannot be well characterized without large time allocation.

Time-resolved measurements of volatile species address key questions in cometary science and could also characterize new classes of interstellar objects[61,62]. *[1] How do the coma and nucleus evolve with heliocentric distance ($R_h$)?* Observations over an entire perihelion passage are critical to understand the evolution of the coma and nucleus. Time-series observations of comets in the UV are particularly lacking, a gap this telescope would fill by observing several cometary passages over the mission lifetime. Nucleus observations at large $R_h$ both pre- and post-perihelion give a broad view of changes in the comet's parent ices. Spatially and temporally resolved observations reveal chemical processes in comae and associate them with active areas on the nucleus. In particular, UV emission spectroscopy and stellar occultations can probe the relationship between $O_2$ and $H_2O$ (Fig. 5)[63]. *Rosetta* measured surprisingly high abundances of $O_2$ in 67P, closely correlated with $H_2O$ throughout perihelion[64]. The telescope could test such correlations as comets rotate and experience outbursts. *[2] What drives outbursts and their frequency, and how often is water ice expelled?* Outbursts have been detected in several comets[65], but the determination of onset and decay as well as characterization of sub-surface, possibly primordial, ice brought into the coma by the phenomenon is difficult to complete without a dedicated telescope with the ability to detect water ice in the UV or the NIR. [3] *What processes dominate in the coma?* Rosetta UV observations have elucidated processes at play in coma chemistry[60]. Atomic oxygen line intensities provide insights into the dominance of electron impact over photodissociation. Similar relationships among other species with emissions in the 100-400 nm range allow a detailed study of coma processes, important for retrieving accurate abundances.

*Science requirements and trades:* For comets, a moving target tracking rate of 216″/hr is recommended[66]. As with Venus, perihelion observations drive sunshield design requirements. Water ice absorptions near 165 nm and 1.5 μm are particularly beneficial. The number of observable comets vs. aperture area and angular/spectral resolution would need to be traded.

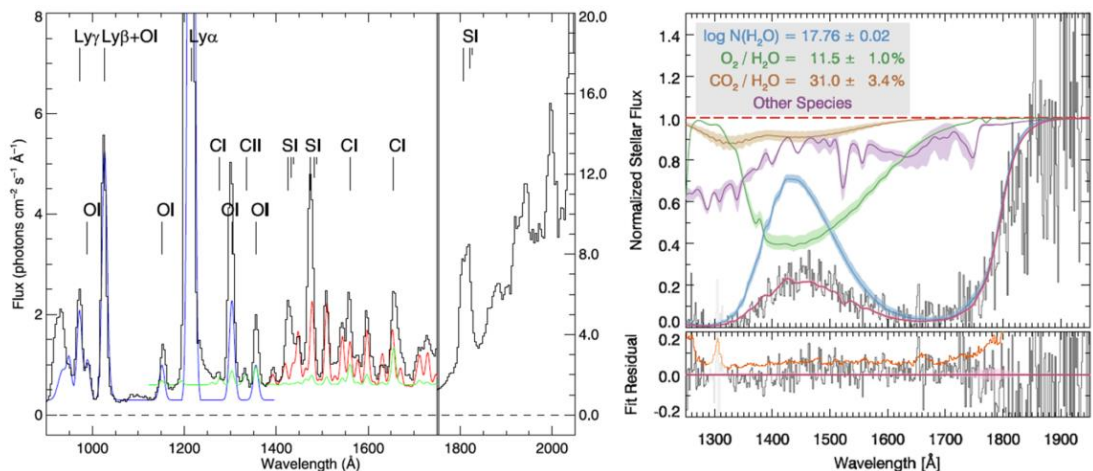

***Fig. 5*** *In multiple comets, the telescope could measure (left) atomic and molecular UV emission to distinguish coma processes such as electron impact (blue, green) and fluorescence (red), and (right) transmission during stellar occultation to determine associations between species such as $O_2$ and $H_2O$, as shown in these examples from Rosetta/Alice data[60,63].*